\documentclass[12pt]{article}
\usepackage{amsmath,latexsym}
\usepackage{graphicx}
\begin{document}

 \title{Vacuumless topological defects in Lyra geometry}
 \author{\Large $F.Rahaman^*$, $S.Mal^*$ and $M. Kalam^\dag$}
\date{}
 \maketitle
 \begin{abstract}
                  Few years ago, Cho and Vilenkin have proposed that topological
                  defects can arise in
                   symmetry breaking models without having
                   degenerate vacua. These types of defects are
                   known as vacuumless defects.
                    In the present work, the gravitational field of a vacuumless
                  global  string and global monopole have been investigated in the context
                  of Lyra geometry.
                  We find the metric of the vacuumless global string and global monopole
                   in the
                  weak field approximations. It has been shown that the
                  vacuumless global string can have repulsive whereas global
                  monopole  exerts
                  attractive gravitational effects on a test
                  particle. It is dissimilar to the case studied in general relativity.
  \end{abstract}

  \footnotetext{Pacs Nos:  04.20 Gz, 04.50 + h, 04.20 Jb   \\
 Key words:  Vacuumless topological defects, Lyra Geometry, Test Particle\\
 $^*$Dept.of Mathematics, Jadavpur University, Kolkata-700 032, India

                                  E-Mail:farook\_rahaman@yahoo.com\\
$^\dag$Dept. of Phys. , Netaji Nagar College for Women, Regent Estate, Kolkata-700092, India.\\
E-Mail:mehedikalam@yahoo.co.in\\
}
    \mbox{} \hspace{.2in}

\title{ \underline{\textbf{Introduction}}: }

Topological defects such as domain walls, monopoles and
    cosmic strings could be produced at a phase transition in the early Universe [1].
    Their nature depends on the symmetry breaking in the field theory under consideration.
    In Cosmology, these defects have been put forward as a possible source for the density
    perturbations which seeded the galaxy formation. \\
   The Lagrangian of a typical symmetry breaking model is of the
   form

    \begin{equation}
              L=\frac{1}{2}\partial_{\mu}\Phi^{a}\partial^{\mu}\Phi^{a}-
              V(f)
         \label{Eq1}
          \end{equation}
     where $\Phi^{a}$ is a set of scalar fields, $a=1,2,......,N
     ,f=\sqrt{(\Phi^{a}\Phi^{a}) }$ and $V(f)$has a minimum at a non
     zero value of $f$.

\pagebreak

      The model has $0(N)$ symmetry and admits
     domain wall, string and monopole solutions for $N=1,2$ and 3
     respectively.\\
     In contrary to the classical idea,  Cho and Vilenkin(CV) [2,3]
     have argued
     that topological defects can also be formed in the models
     where $ V(f)$ is maximum at $f=0$ and it decreases monotonically
     to zero for $f \longrightarrow\infty$  without having any
     minima.\\
     They have provided an example of the above idea as

     \begin{equation}
              V(f)= \lambda M^{4+n}(M^{n}+f^{n})^{-1}
         \label{Eq2}
          \end{equation}

     where M, $\lambda$ and n are positive constants.

      In non perturbative super string models, this type of potential can be found frequently.
      Defects
      arising in this model are termed as vacuumless.
Recent observations of the luminosity-redshift relation of type Ia
supernovae suggest that the Universe is accelerating at the
present epoch. Physicists are trying to search for a matter field
which is responsible for this accelerating expansion of the
Universe.
   This matter field is called "Quintessence" or Q-matter. It is
   readily understand that this current cosmological state of the
   Universe requires Q-matter having  a scalar
field with a potential which generates a sufficient negative
pressure at the present epoch. Examples of Q-matter are
fundamental fields or macroscopic objects and notion of vacuumless
defects can be used to explain this astonishing phenomena
theoretically as scalar field with potential like (2) can act as
Quintessence models [4].

      CV have studied the gravitational field of topological defects in the
      above models within the frame work of general relativity[3]. But at sufficiently
      high energy scales, it seems likely that gravity is not given by Einstein's
      action and also Einstein's general theory of relativity
      could not able to explain the acceleration of the Universe. The
      Big Bang singularity is another drawbacks of Einstein's
      theory. So alternating theories are proposed time to time.

      In last few decades there has been considerable interest in alternative
      theories of gravitation. The most important among them being scalar-tensor theories
      proposed by Lyra [5] and Brans-Dicke [6]. Lyra suggested a modification of
      Riemannian geometry which may also be considered as a modification of Weyl's geometry.
      In Lyra's geometry, Weyl's concept of gauge, which is essentially a metrical concept,
      is modified by the introduction of a gauge function into the structure less
      manifold.\\
This alternating theory is of interest because it produces effects
similar to those produced in Einstein's
      theory. Also vector field in this theory plays similar role
      to cosmological constant in general relativity.

\pagebreak

     In Lyra's geometry, Einstein's
      field equations are [7]
      \\

      \begin{equation}
              R_{ab} - \frac{1}{2}g_{ab}R + \frac{3}{2}\phi_a
              \phi_b - \frac{3}{4}g_{ab}\phi_c\phi^c = - 8\pi G
              T_{ab}
           \end{equation}

          where $\phi_a$ is the displacement vector and other symbols have their usual meaning
    as in Riemannian geometry.\\
    Subsequent investigations were done by several authors in scalar tensor theory and
    cosmology as well as topological defects  within the framework of Lyra geometry [8-21].\\
    In  recent papers, Sen [22] and Rahaman [23] have studied the gravitational field of vacuumless
   topological defects in Brans-Dicke theory.\\
    In this work, we have study the gravitational field of vacuumless topological defects
    in Lyra geometry with constant displacement vector $ \phi_a =(\beta_0,0,0,0)$
    under the weak field approximation of the field equations.
  We also analyze  the gravitational effects on test particles.

$ $

 { 2. \textbf{The Basic Equations for constructing vacuum less
cosmic string: }}

          According to CV [2,3] a vacuumless cosmic string is described by a scalar
          doublet $ ( \Phi_1 , \Phi_2 ) $ with a power law potential (2). In cylindrical
          coordinates $( r,\theta,z) $, one can assume the ansatz as $ \Phi_1 = f(r)cos\theta,
          \Phi_2 = f(r)sin\theta.$
 For vacuum less global string the flat space time solution for $ f(r)$ is given
 by

 \begin{equation}
              f(r)= aM(\frac{r}{\delta})^{\frac{2}{n+2}}
         \label{Eq15}
          \end{equation}

     where $\delta=\frac{1}{M\surd\lambda}$ is the core radius of the string , r is the
     distance from the string axis and
     $a=(n+2)^{\frac{2}{n+2}}(n+4)^{-\frac{1}{n+2}}$.\\
     The solution (4) applies for

     \begin{equation}
               \delta \ll r \ll R
         \label{Eq16}
          \end{equation}

    where R is the cut off radius determined by the nearest string .\\
    For a vacuumless cosmic string the space time is static, cylindrically symmetric
    and also has a symmetry with respect to Lorenz boost along the string axis.
    One can write the corresponding line element as

    \begin{equation}
               ds^2=A(r) ( - dt^2 + dz^2)+A(r)dr^2+r^2B(r)d\theta^2
         \label{Eq17}
          \end{equation}

The general energy momentum tensor for the vacuumless string is
given by
       \begin{equation}
               T_t^t=T_z^z=\frac{(f^\prime)^2}{2A}+\frac{f^2(1-\alpha)^2}{2Br^2}+
               \frac{(\alpha^\prime)^2}{2e^2Br^2}+ V(f)
         \label{Eq18}
    \end{equation}
    \begin{equation}
            T_r^r=-\frac{(f^\prime)^2}{2A}+\frac{f^2(1-\alpha)^2}{2Br^2}-
            \frac{(\alpha^\prime)^2}{2e^2Br^2} + V(f)
          \label{Eq19}
    \end{equation}
    \begin{equation}
               T_\theta^\theta= \frac{(f^\prime)^2}{2A}-\frac{f^2(1-\alpha)^2}{2Br^2}-
            \frac{(\alpha^\prime)^2}{2e^2Br^2} + V(f)
           \label{Eq20}
     \end{equation}

     where string ansatz for the gauge field is $A_\theta(r)=-\frac{ \alpha (r)}{er }$.\\
    $T_a^b$ 's with $\alpha = 0$ are that for global string.\\

    The field equations (3) for the metric (6) are
    \begin{equation}
               \frac{1}{2}\frac{A^\prime B^\prime}{A^2 B} +
               \frac{(A^\prime)^2}{4A^3} + \frac{A^\prime}{r A^2}+
               \frac{3}{4} \frac{\beta_0^2}{A} = 8 \pi G T_r^r
           \label{Eq21}
     \end{equation}
    \begin{equation}
              - \frac{3}{4}\frac{(A^\prime)^2}{A^3} +
               \frac{A^{\prime\prime}}{A^2} +
               \frac{3}{4} \frac{\beta_0^2}{A} = 8 \pi G T_\theta^\theta
           \label{Eq22}
     \end{equation}
     \begin{equation}
              - \frac{1}{2}\frac{(A^\prime)^2}{A^3} +\frac{1}{2}
               \frac{A^{\prime\prime}}{A^2} +\frac{1}{2}
               \frac{B^{\prime\prime}}{BA}+\frac{B^\prime}{rAB}
               -  \frac{3}{4}\frac{(B^\prime)^2}{AB^2}
               - \frac{3}{4} \frac{\beta_0^2}{A} = 8 \pi G T_t^t
           \label{Eq23}
     \end{equation}

     ['$^\prime$' indicates differentiation w.r.t. r ]

 {  3. \textbf{Solutions in the weak field approximations }:}

    Now we will consider the solution of a vacuumless cosmic string in the weak field approximation. At this stage,
   let us assume that \\
    \begin{equation}
        A(r) = 1+\beta(r)  ,  B(r)= 1+\gamma(r)
    \label{Eq24}
    \end{equation}
        where $\beta , \gamma \ll 1$ . For global vacuum less string ,
        one can use the flat space approximation for $f(r)$ in $eq.(4)$ for $r \gg \delta $
        and the form of $ V(f)$ given in $eq.(2)$.\\
        Now under these weak field approximations, the field equations
        take the following forms:
        \begin{equation}
              \frac{\beta^\prime}{r} + \frac{3}{4} \beta_0^2 =
              \frac{D(n^2 + 6n +16)}{(n+2)^2} r^{-b}
           \label{Eq25}
        \end{equation}
        \begin{equation}
              \beta^{\prime\prime} + \frac{3}{4} \beta_0^2 =
              \frac{D(n-4)}{(n+2)} r^{-b}
           \label{Eq26}
        \end{equation}
        \begin{equation}
              \frac{1}{2}\beta^{\prime\prime} +
              \frac{1}{2}\gamma^{\prime\prime}+
              \frac{\gamma^\prime}{r}- \frac{3}{4} \beta_0^2 =
              \frac{D(n+4)}{(n+2)} r^{-b}
           \label{Eq27}
        \end{equation}
     where $ D = 8\pi G a^2 M^2 \delta ^\frac{- 4}{(n+2)}$ and $ b =
     \frac{2n}{(n+2)}.$

     From $ eq.(14)$ , we get the following solution of $ \beta $ as
     \begin{equation}
              \beta= - \frac{3}{2} {\beta_0^2 r^2}-
              \frac{D(n^2 + 6n +16)}{(2-b)(n+2)^2} r^{-b+2}
           \label{Eq28}
        \end{equation}
      Also from $ eq.(15)$ , we get the following solution of $ \beta $ as
      \begin{equation}
              \beta= - \frac{3}{2} {\beta_0^2 r^2}-
              \frac{D(n-4)}{(1-b)(2-b)(n+2)} r^{-b+2}
           \label{Eq29}
        \end{equation}
        For consistency, we must have that the second term of both the equations is same i.e.
        $n$ has the value as  $n = 2.718608172.$\\
        Also we get the solution of $\gamma$ as
        \begin{equation}
              \gamma= \frac{1}{8} {\beta_0^2 r^2}+
              \frac{D(3n+4)}{(3-b)(2-b)(n+2)} r^{-b+2}
           \label{Eq30}
        \end{equation}
\begin{figure}[htbp]
    \centering
        \includegraphics[scale=.3]{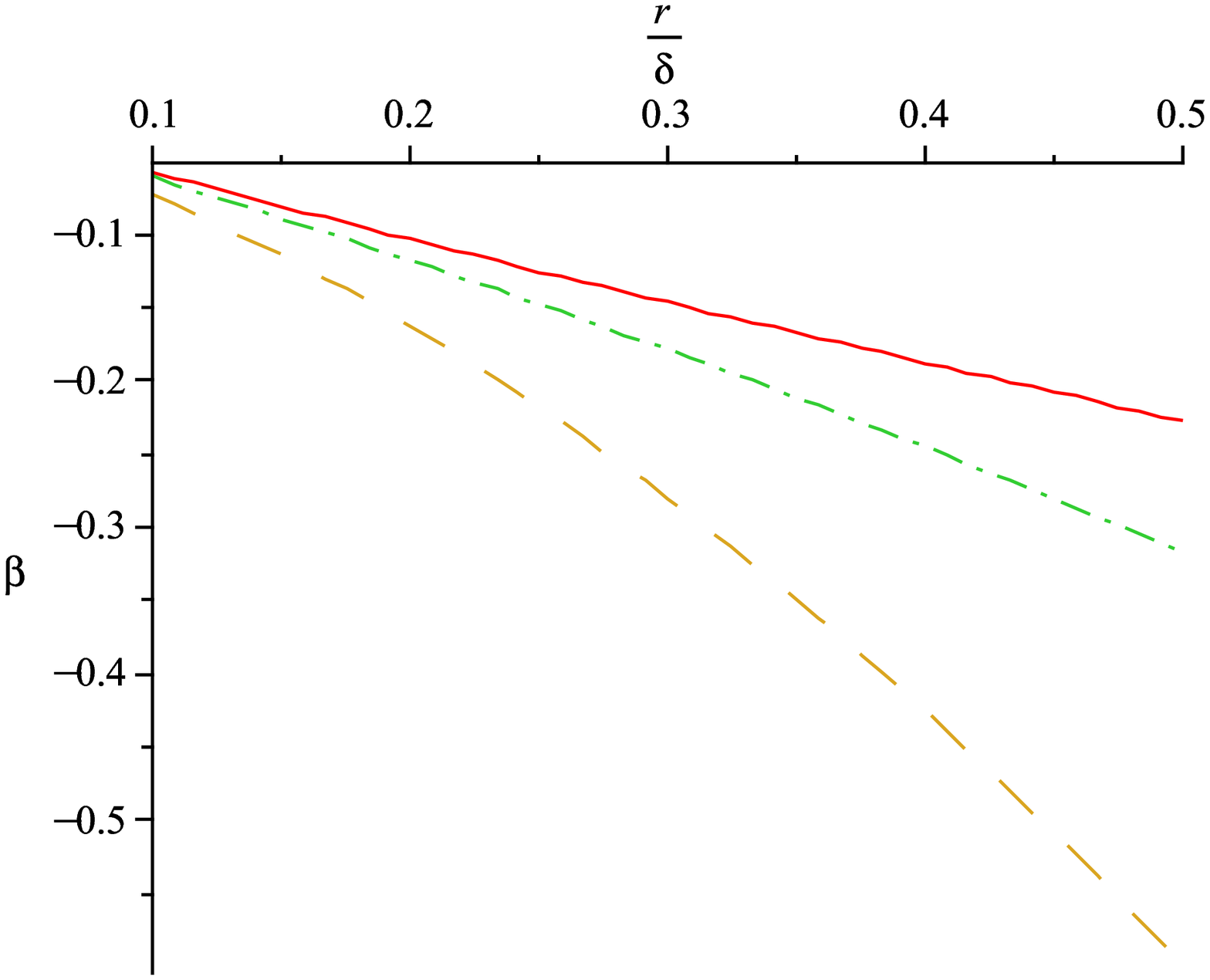}
    \caption{We show the variation of $\beta $ with respect to $\frac{r}{\delta}$  for different values of displacement vector
    $\beta_0  $ and choosing other parameters as constants ($\beta_0 = .1$, for red line; $\beta_0 = .5$,
    for green line; $\beta_0 = 1$, for yellow line ).}
    \label{fig:cs}
\end{figure}
\begin{figure}[htbp]
    \centering
        \includegraphics[scale=.3]{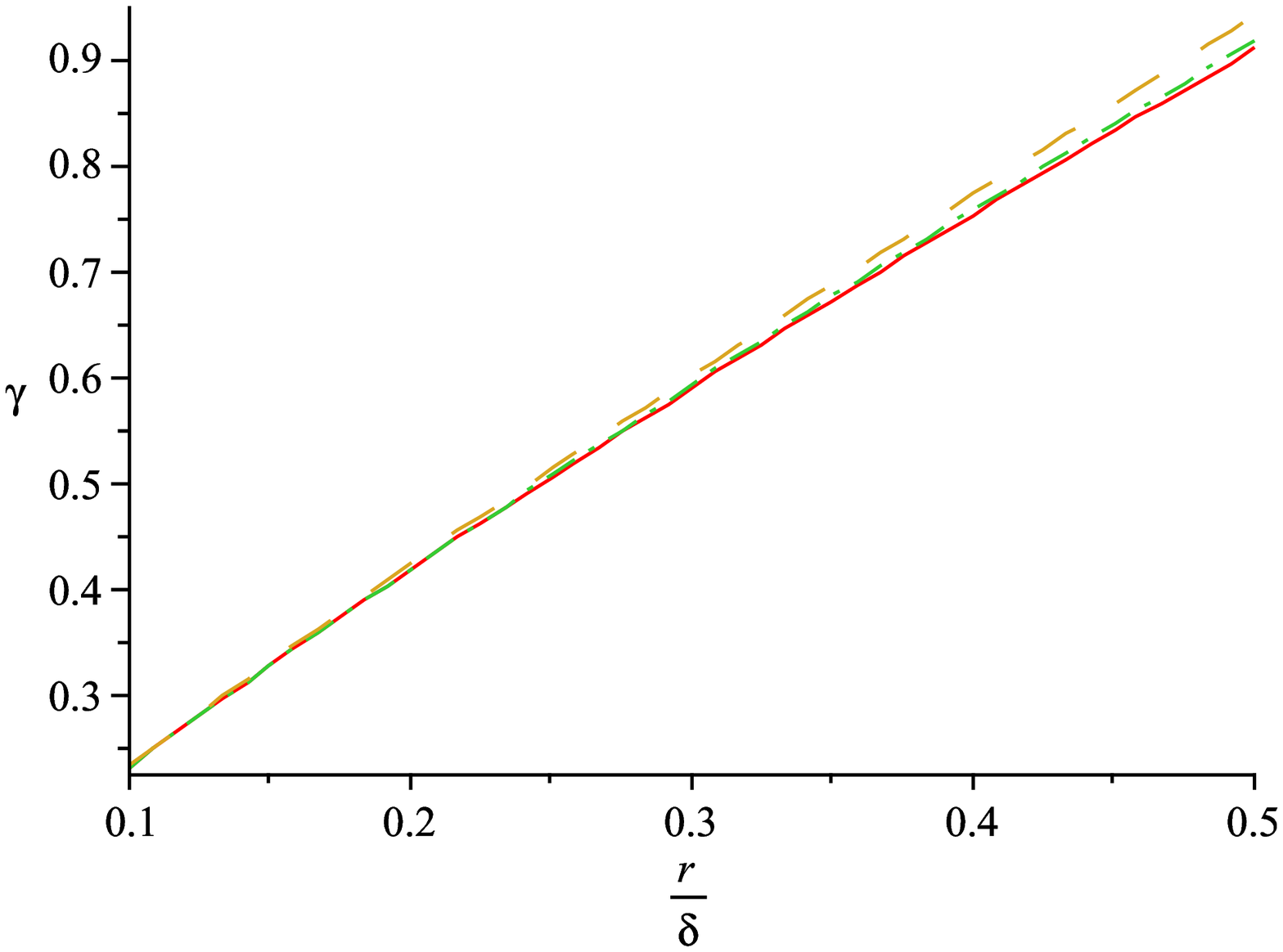}
    \caption{We show the variation of $\gamma $ with respect to $\frac{r}{\delta}$ for different values of displacement vector
    $\beta_0  $ and choosing other parameters as constants ($\beta_0 = .1$, for red line; $\beta_0 = .5$, for green line; $\beta_0 = 1$, for yellow line ).  }
    \label{fig:cs}
\end{figure}

Thus the solution for vacuumless cosmic string in Lyra geometry
will be taken the following form as

           $     ds^2=\left[1 - \frac{3}{2} {\beta_0^2 r^2}-
              \frac{D(n-4)}{(1-b)(2-b)(n+2)} r^{-b+2} \right]( - dt^2 + dz^2 + dr^2)+r^2\left[1 + \frac{1}{8} {\beta_0^2 r^2}+
              \frac{D(3n+4)}{(3-b)(2-b)(n+2)}
              r^{-b+2}\right]d\theta^2$
\begin{equation}
         \label{Eq5}
          \end{equation}
\begin{figure}[htbp]
    \centering
        \includegraphics[scale=.3]{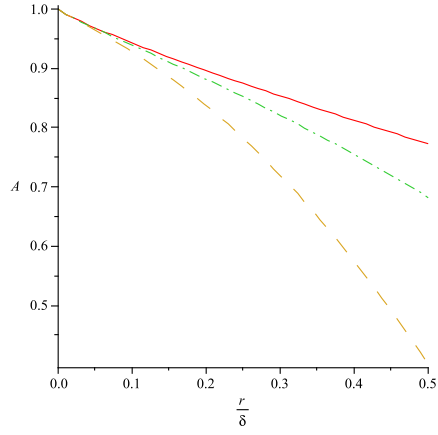}
    \caption{We show the variation of $A \equiv g_{tt}\equiv g_{zz}\equiv g_{rr} $ with respect to $\frac{r}{\delta}$
    for different values of displacement vector
    $\beta_0  $ and choosing other parameters as constants ($\beta_0 = .1$, for
     red line; $\beta_0 = .5$, for green line; $\beta_0 = 1$, for yellow line ).  }
    \label{fig:cs}
\end{figure}
\begin{figure}[htbp]
    \centering
        \includegraphics[scale=.35]{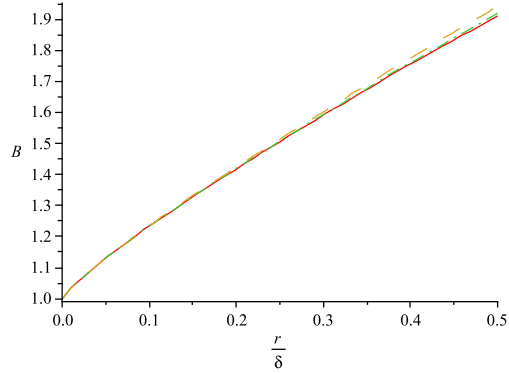}
    \caption{We show the variation of $B \equiv g_{\theta \theta} $ with respect to $\frac{r}{\delta}$
     for different values of displacement vector
    $\beta_0  $ and choosing other parameters as constants ($\beta_0 = .1$, for red
     line; $\beta_0 = .5$, for green line; $\beta_0 = 1$, for yellow line ).  }
    \label{fig:cs}
\end{figure}

\pagebreak

{\textbf{ 4. Gravitational effects on test particles }:}

        The repulsive and attractive character of the global string can be discussed
        by either studying the time like geodesics in the space time or analyzing the
        acceleration of an observer who is at rest relative to the
        string.\\
        We now calculate the radial acceleration vector $A^r$ of a
        particle that remains stationary ( i.e. $ V^1= V^2= V^3= 0 $ ) in the field of
        the string .\\
        Let us consider an observer with four velocity $ V_i = \sqrt{g_{00}} \delta_i^t $
        .\\
        Now $ A^r = V^1_{; 0}V^0 = \Gamma^1_{00}V^0 V^0 $.

        So for our space time (20),

        \begin{equation}
              A^r = - 3 {\beta_0^2 r}+\frac{D(-n+4)}{(1-b)(n+2)(1 + \beta)^3} r^{-b+1}
           \label{Eq32}
        \end{equation}

        We see that for $ n = 2.718608172 $ the expression $A^r$ is  negative
         and in this
        case gravitational forces varies with the radial distance.
        This indicates that particle accelerates towards the
        string in the radial direction in order to keep it rest.
        This implies that the string has a repulsive influence on
        the test particle [24].\\

\begin{figure}[htbp]
    \centering
        \includegraphics[scale=.4]{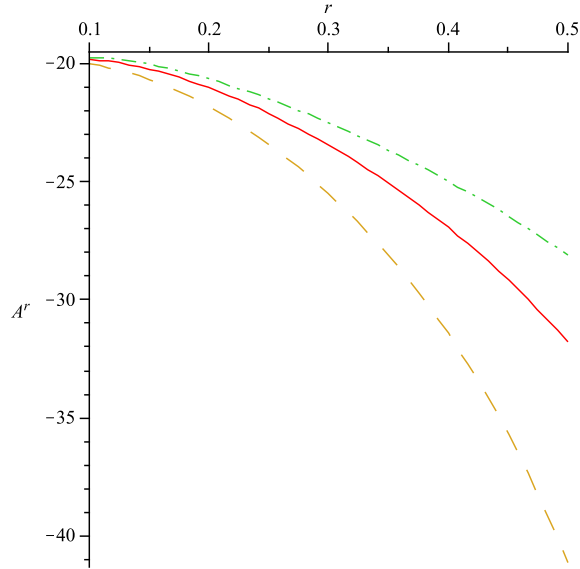}
    \caption{We show the variation of acceleration $(A^r)$ with respect to r for different values of displacement vector
    $\beta_0  $ and choosing other parameters as constants ($\beta_0 = .3$, for red line; $\beta_0 = .1$, for green line; $\beta_0 = .5$, for yellow line ).  }
    \label{fig:cs}
\end{figure}

\pagebreak

{ 5. \textbf{The Basic Equations for constructing vacuum less
global monopole: }}

According to Bariolla and Vilenkin [25], the monopole is
associated with
 a triplet of scalar fields
    $\Phi^{a},a=1,2,3.$ and the monopole ansatz is taken as
    $\Phi^{a}=f(r)\frac{x^{a}}{r}$, where r is the distance from the monopole center.
    Cho and Vilenkin  have shown that for the power law potential (2),
    the field equation for $f(r)$ admits a solution of
 the form[2,3]

 \begin{equation}
              f(r)= aM(\frac{r}{\delta})^{\frac{2}{n+2}}
         \label{Eq3}
          \end{equation}
     where $\delta=\frac{1}{M\surd\lambda}$ is the core radius of the monopole , r is the
     distance from the monopole center and
     $a=(n+2)^{\frac{2}{n+2}}(n+4)^{-\frac{1}{n+2}}$.\\

     It is argued that the solution (22) can be applied in the region for
     \begin{equation}
               \delta \ll r \ll R
         \label{Eq4}
          \end{equation}
    where the cut off radius R is set by the distance to the
    nearest anti monopole .\\

    For a vacuum less monopole the space time is static
    , spherically symmetric and  the corresponding line
    element can be taken as

    \begin{equation}
                ds^2=   e^{\nu(r)} dt^2 - e^{\lambda(r)} dr^2 - r^2( d\theta^2+sin^2\theta
               d\phi^2)
         \label{Eq5}
          \end{equation}

    The components of the stress energy  tensors for the vacuumless
    monopole are given by

    \begin{equation}
               T_t^t=\frac{(f^\prime)^2}{2e^{\lambda(r)}}+\frac{f^2w^2}{r^2}+\frac{1}{2e^2r^2}\left[\frac{(w^\prime)^2}{e^{\lambda(r)}}
               +\frac{(1-w^2)^2}{2r^2}\right]+ V(f)
         \label{Eq6}
    \end{equation}
    \begin{equation}
            T_r^r=-\frac{(f^\prime)^2}{2e^{\lambda(r)}}+\frac{f^2w^2}{r^2}+\frac{1}{2e^2r^2}\left[\frac{(w^\prime)^2}{e^{\lambda(r)}}
               +\frac{(1-w^2)^2}{2r^2}\right]+ V(f)
          \label{Eq7}
    \end{equation}
    \begin{equation}
               T_\theta^\theta=T_\phi^\phi=\frac{(f^\prime)^2}{2e^{\lambda(r)}}+\frac{1}{2e^2r^2}\left[\frac{(w^\prime)^2}{e^{\lambda(r)}}
               +\frac{(1-w^2)^2}{2r^2}\right]+ V(f)
           \label{Eq8}
     \end{equation}

     For monopole, the gauge field is $  A_i^a(r)=\frac{[1-w(r)]\epsilon^{aij}x^j}{er^2}$ .
    $T_a^b$ 's with $w =1$ are that for global monopole. For global
    vacuum less monopole, one can use the flat space approximation
    for $f(r)$ in (22)  as well as $V(f)$ given in in (2) for
    $r\gg\delta$.\\This follows from the fact that in this case
    the gravity would not much influence on monopole structure.

\pagebreak

   Here the field equations are
\begin{equation}e^{-\lambda}
\left[-\frac{\lambda^\prime}{r} + \frac{1}{r^2}
\right]-\frac{1}{r^2} + \frac{3}{4}e^{-\nu}\beta_0^2 = 8\pi G C
r^{-b}
\end{equation}
\begin{equation}e^{-\lambda}
\left[\frac{1}{r^2}+\frac{\nu^\prime}{r}\right]-\frac{1}{r^2} -
\frac{3}{4}e^{-\nu}\beta_0^2 = 8\pi GD r^{-b}
\end{equation}
\begin{equation}\frac{1}{2} e^{-\lambda}
\left[\frac{1}{2}(\nu^\prime)^2+ \nu^{\prime\prime}
-\frac{1}{2}\lambda^\prime\nu^\prime + \frac{1}{r}({\nu^\prime-
\lambda^\prime})\right] + \frac{3}{4}e^{-\nu}\beta_0^2 = 8\pi G E
r^{-b}
\end{equation}

             where \\

        $
                        C = \left[ \frac{2 a^2M^2}{2(n+2)} + a^2M^2+ \frac{M^2}{
                        a^2}\right]
                        \delta^\frac{-4}{n+2}  ,\\
                       D = \left[- \frac{2 a^2M^2}{2(n+2)} + a^2M^2+ \frac{M^2}{
                        a^2}\right]
                        \delta^\frac{-4}{n+2}  ,\\
                        E = \left[ \frac{2 a^2M^2}{2(n+2)} +  \frac{M^2}{
                        a^2}\right]
                        \delta^\frac{-4}{n+2}  ,\\
                         b = \frac{2n}{(n+2)}   .$

 {\textbf{6. Solutions in the weak field approximations :}}

   Now we will consider the solution of a vacuumless monopole in the weak field approximation. At this stage,
   let us assume that

        $e^{\nu(r)} = 1+f(r), e^{\lambda(r)}= 1+g(r).$

     where $ f,g \ll 1 $.\\
     In this approximations $eqs.(9)-(12)$ take the following
     forms as
     \begin{equation}
               \frac{g}{r^2} +\frac{g^\prime}{r} -\frac{3}{4}\beta_0^2 = - 8\pi G C  r^{-b}
           \label{Eq14}
     \end{equation}
      \begin{equation}
              - \frac{g}{r^2} +\frac{f^\prime}{r} -\frac{3}{4}\beta_0^2 =  8\pi G D  r^{-b}
           \label{Eq14}
     \end{equation}
     \begin{equation}
              f^{\prime\prime} + \frac{(f^\prime - g^\prime)}{r} -\frac{3}{4}\beta_0^2 =8\pi G E  r^{-b}
           \label{Eq16}
     \end{equation}
     Solving these equations , we get\\
    \begin{equation}
           g = \frac{1}{4}r^2\beta_0^2 - \left[ \frac{8\pi G C }{(3-b)} \right] r^{2-b}
           \label{Eq18}
     \end{equation}
    \begin{equation}
           f = \frac{1}{2}r^2\beta_0^2 - \left[ \frac{8\pi G  }{(2-b)} \right]
           \left[ \frac{-D+C  }{(3-b)} \right] r^{2-b}
           \label{Eq19}
     \end{equation}

\begin{figure}[htbp]
    \centering
        \includegraphics[scale=.4]{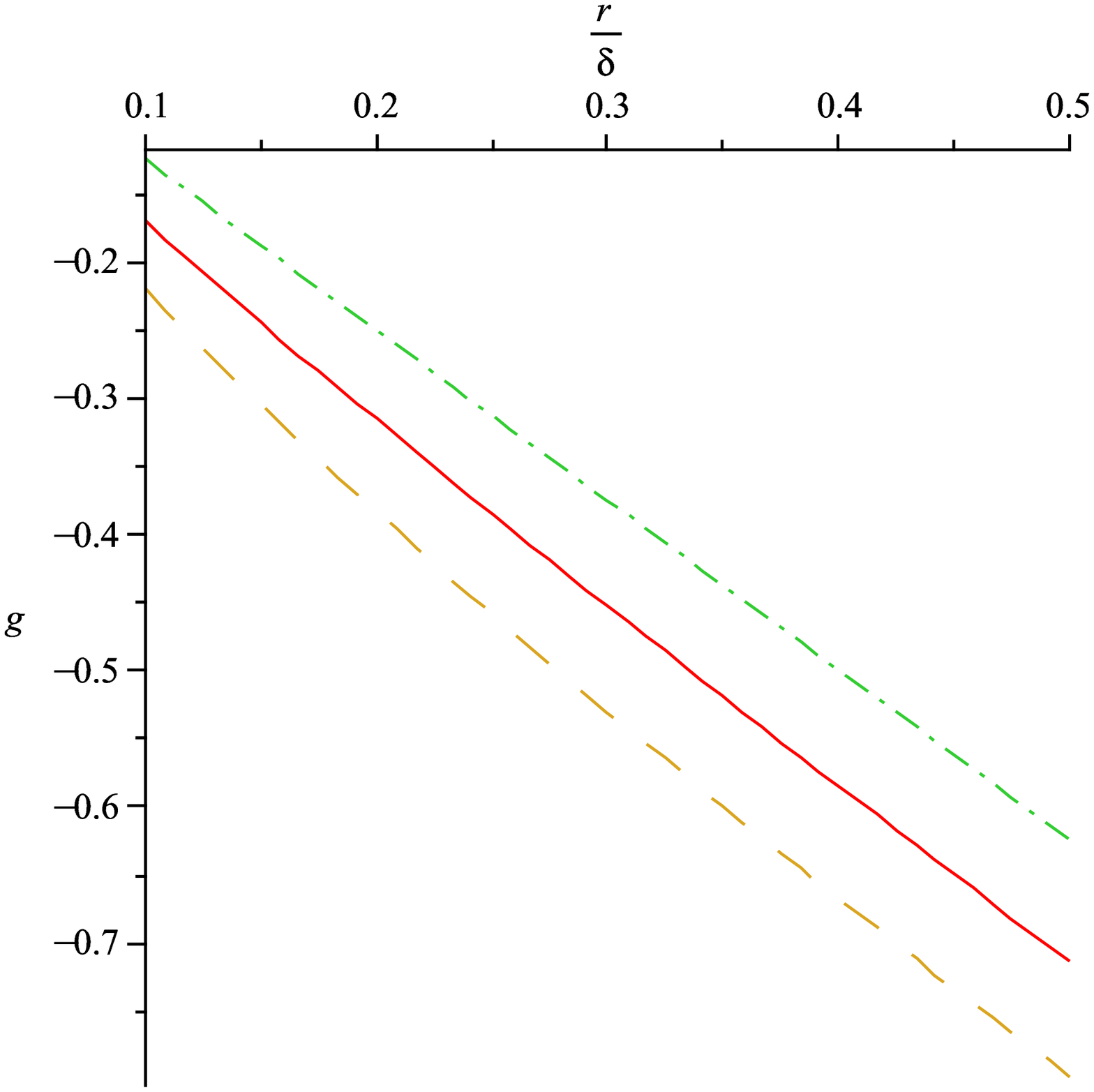}
    \caption{We show the variation of $g$ with respect to $\frac{r}{\delta}$ for different values of
    n and choosing other parameters as constants ($n = 2.5$, for red line; $n = 2$,
     for green line; $n= 3$, for yellow line ).
     }
    \label{fig:mono}
\end{figure}
\begin{figure}[htbp]
    \centering
        \includegraphics[scale=.4]{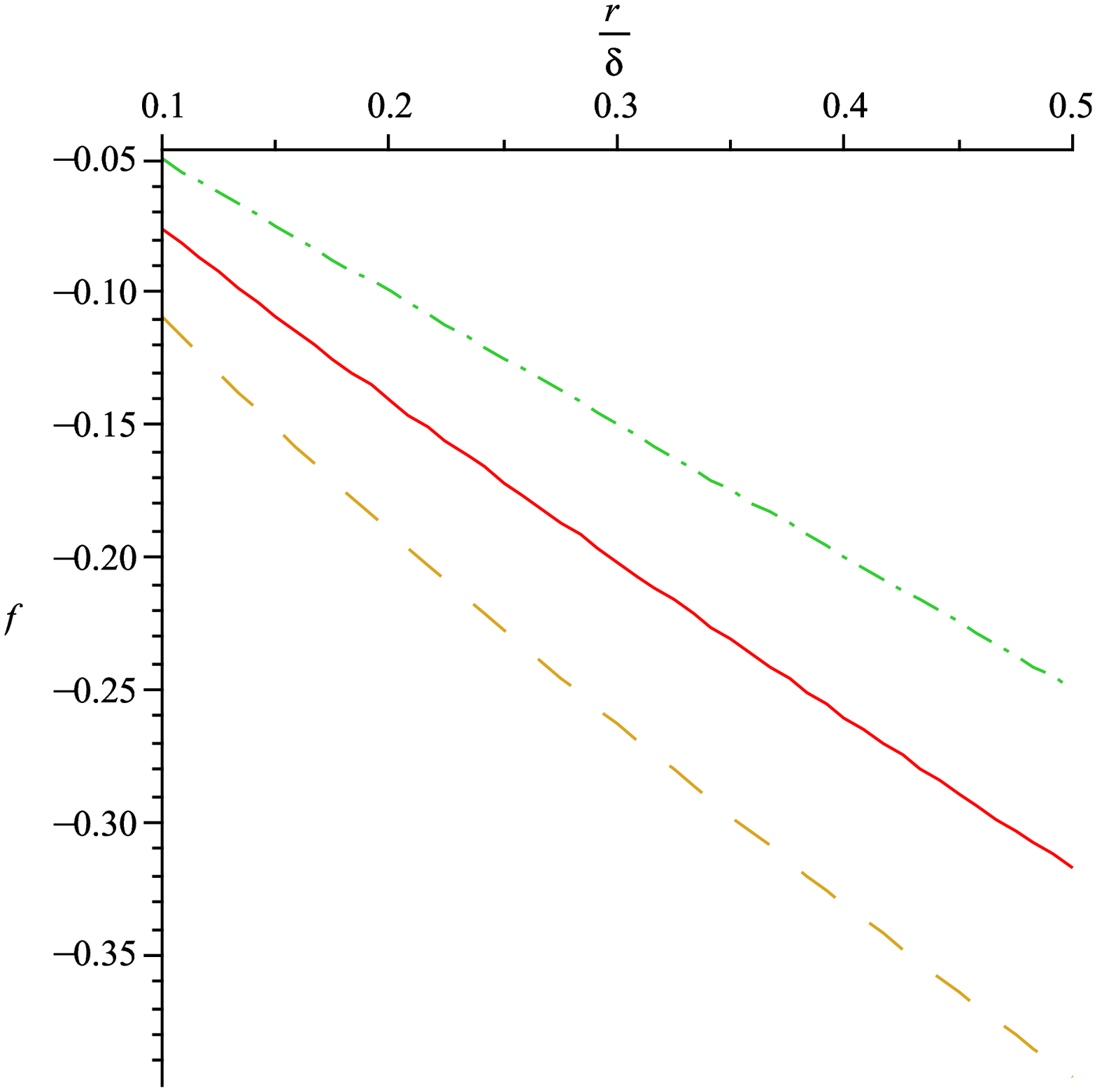}
    \caption{We show the variation of $f $ with respect to $\frac{r}{\delta}$  for different values
    of n and choosing other parameters as constants ($n = 2.5$, for red line; $n = 2$,
     for green line; $n= 3$, for yellow line ).
     }
    \label{fig:mono}
\end{figure}

Thus the solution for vacuumless global monopole in Lyra geometry
will be  taken the following form as
\begin{equation}
                ds^2=  \left [1+ \frac{1}{2}r^2\beta_0^2 + \pi G M^2 K(n)  \left(\frac{r}{\delta}\right)^{2-b}\right]dt^2 -
                 \left [1+ \frac{1}{4}r^2\beta_0^2 + \pi G M^2 C(n)  \left(\frac{r}{\delta}\right)^{2-b}\right] dr^2 - r^2 d\Omega_2^2
         \label{Eq5}
          \end{equation}
where $K(n) $ and $C(n)$ are complicated functions of n. If we put
$\beta_0 = 0$ i.e. in the absence of the displacement vector our
metric transforms to
\begin{equation}
                ds^2=  \left [1 + \pi G M^2 K(n)  \left(\frac{r}{\delta}\right)^{2-b}\right]dt^2 -
                 \left [1+  \pi G M^2 C(n)  \left(\frac{r}{\delta}\right)^{2-b}\right] dr^2 - r^2 d\Omega_2^2
         \label{Eq5}
          \end{equation}
          which is same form of $CV^{'s}$ solution in general
          relativity.

\begin{figure}[htbp]
    \centering
        \includegraphics[scale=.4]{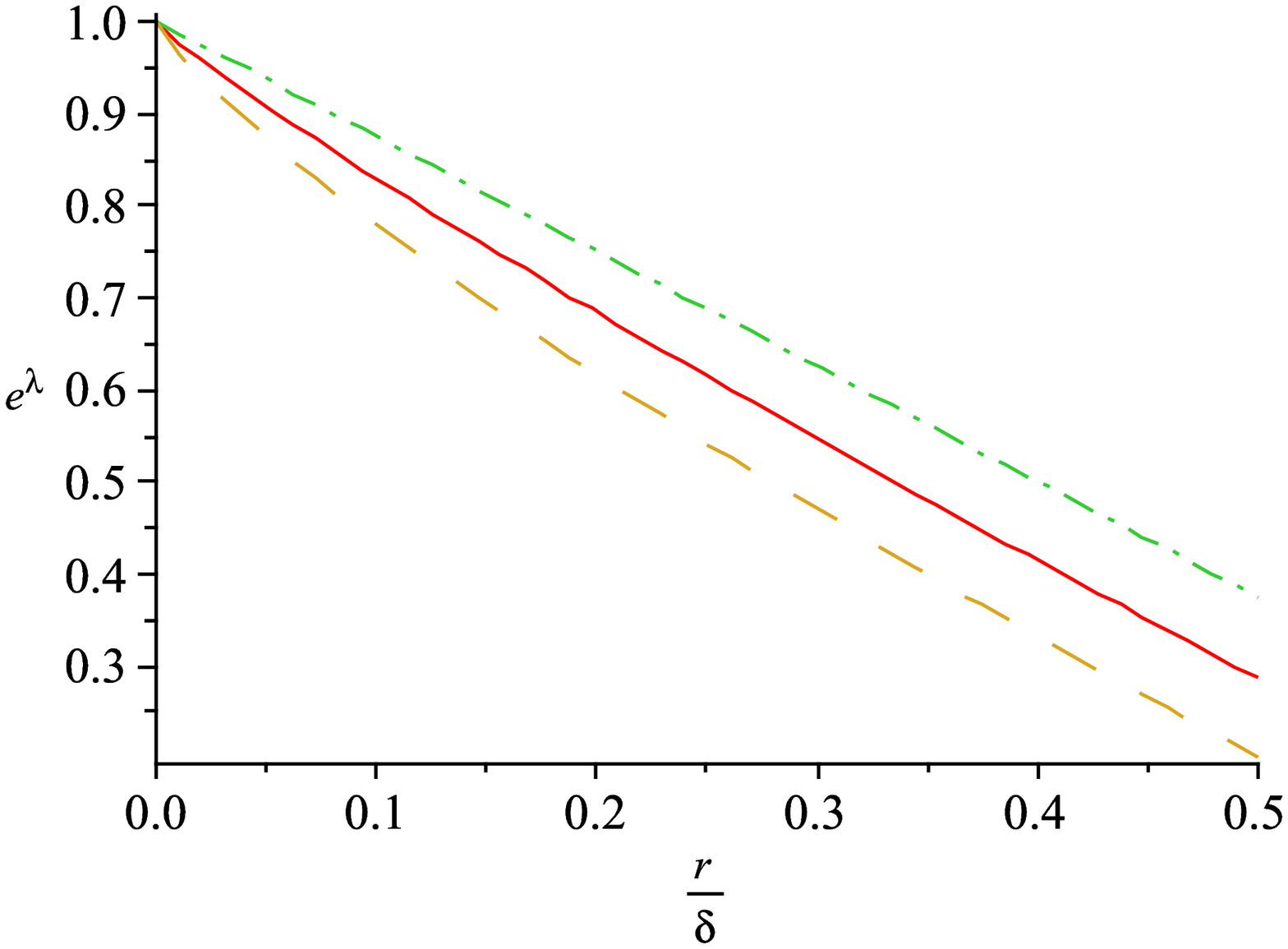}
    \caption{We show the variation of $e^\lambda $ with respect to $\frac{r}{\delta}$ for different values
    of n  and choosing other parameters as constants ($n = 2.5$, for red line; $n = 2$,
     for green line; $n= 3$, for yellow line ).
     }
    \label{fig:mono}
\end{figure}
\begin{figure}[htbp]
    \centering
        \includegraphics[scale=.35]{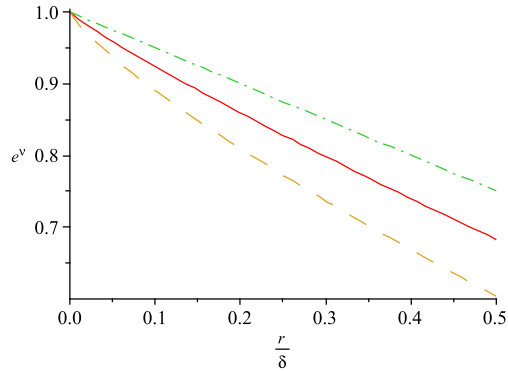}
    \caption{We show the variation of $e^\nu $ with respect to $\frac{r}{\delta}$ for different values
    of n and choosing other parameters as constants ($n = 2.5$, for red line; $n = 2$,
     for green line; $n= 3$, for yellow line ).
     }
    \label{fig:mono}
\end{figure}

\pagebreak

     {\textbf{7. Gravitational effects on test particles  :}}

    Let us consider a relativistic particle of mass m,moving in
    the gravitational field of monopole described by equation (36)
    .\\
    According  to the formalism of Hamilton and
    Jacobi (H-J), the H-J equation is [26]
    \begin{equation}
          \frac{1}{B(r)}(\frac{\partial S}{\partial t})^2 -  \frac{1}{A(r)}(\frac{\partial S}{\partial r})^2
          -  \frac{1}{r^2}(\frac{\partial S}{\partial \theta})^2
          -  \frac{1}{r^2 \sin ^2\theta}(\frac{\partial S}{\partial
          \phi})^2 + m^2 = 0
           \label{Eq21}
     \end{equation}
     where

     $
     B(r) = \left [1+ \frac{1}{2}r^2\beta_0^2 + \pi G M^2 K(n)
     \left(\frac{r}{\delta}\right)^{2-b}\right]$
     and
     $A(r) = \left [1+ \frac{1}{4}r^2\beta_0^2 + \pi G M^2 C(n)  \left(\frac{r}{\delta}\right)^{2-b}\right]$

     In order to solve the particle differential equation , let us
     use the separation of variables for the H-J function S as
     follows [26] .\\
     \begin{equation}
           S(t,r,\theta,\phi)=-Et + S_1(r) + S_2(\theta) + J\phi
           \label{Eq22}
     \end{equation}
     Here the constants E and J are identified as the energy and
     angular momentum of the particle .\\
      The radial velocity of the particle is( For
      detail calculations see reference (26) ).
    \begin{equation}
           \frac{dr}{dt} = \frac{B}{E\sqrt{A}}\sqrt{\frac{E^2}{B}
           + m^2 - \frac{p^2}{r^2}}
           \label{Eq23}
     \end{equation}
     where $ p$ is the separation constant .

      The turning points of
     the trajectory are given by $ \frac{dr}{dt}= 0$ and as a
     consequence the potential curves are \\
    \begin{equation}
           \frac{E}{m} = \sqrt{\left [1+ \frac{1}{2}r^2\beta_0^2 + \pi G M^2 K(n)  \left(\frac{r}{\delta}\right)^{2-b}\right]
           \left[\frac{p^2}{m^2r^2}-1\right]}\equiv V
            \label{Eq24}
     \end{equation}\\
     In this case the extremals of the potential curve are the
     solutions of the equation \\
     \begin{equation}
           m^2 H (2-b)r^{4-b} + bHr^{2-b} + m^2r^4\beta_0^2- 2 p^2 = 0
           \label{Eq25}
     \end{equation}\\
     where $ H = \pi G M^2 K(n)
     \delta^{b-2}$.

     This equation has at least one positive real root provided
     (-b+2) is an positive integer. So it is possible to have
     bound orbit for the test particle. Thus the gravitational
     field of the vacuumless global monopole is shown to be attractive in
     nature but here we have to imposed some restriction on the
     constant "n". This effect is absent in general relativity case.

\pagebreak

{ \textbf{8. Conclusion :}}

      Observations of the flatness of galactic rotation curves
      indicate the galaxies, cluster of galaxies and
      super clusters are filled out with 90 percent of nonluminous
      matter (dark matter). Recently, it has been suggested that
      some of the topological defects such as monopole could be
      present in the galactic dark matter and these topological defects are responsible for
       the structure formation of the Universe [27-30]. So,
       topological defects
       have been revived in the recent years. This work extends the earlier work by Cho and Vilenkin regarding
       the gravitational field of a vacuumless topological defects, namely, cosmic string and global monopole
        to the scalar tensor theory
       based on Lyra geometry. We see that in going from general relativity  to  scalar
       tensor theory based on Lyra geometry both space time curvature and topology are
       affected by the present of the displacement vector. Our study of the motion of
       the test particle reveals that the vacuumless global string in Lyra geometry exerts
       gravitational force which is repulsive in nature. It is
       similar to the case of a  vacuumless global string in general relativity
       where the vacuumless global string in general relativity has  repulsive
       gravitational effect.\\
Whereas vacuumless global monopole in Lyra geometry exerts
gravitational force which is
                                 attractive in nature. It is dissimilar to the case of a vacuumless global monopole in general relativity. In the absence of
                                  the displacement vector i.e. $\beta_0 = 0$, our solution coincides
                                   with CV solution (with proper choices of the arbitrary
                                   constants). Since our vacuumless
                                   monopole exerts gravitational
                                   pull on surrounding particles,
                                   so it seems important to
                                   consider vacuumless
                                   monopole field as a candidate
                                   for galactic dark matter.


\end{document}